\documentclass{article}

\usepackage{graphicx,color}      
\usepackage{dcolumn}          
\usepackage{bm}                  
\usepackage{float}                

\title{ERBU, Expanding Rubber Band Universe}

\author{Domingos Soares\footnote{\small dsoares@fisica.ufmg.br} \\ Physics Department, 
ICEx, UFMG --- C.P. 702 \\ 30123-970,  Belo Horizonte, Brazil} 

\date{October 06, 2014}

\begin{document}


\def\Ho{$H_\circ$}
\def\omegao{$\Omega_\circ$}
\def\to{$t_\circ$}

\maketitle

\begin{abstract}
I put forward a simple unidimensional mechanical analogue of the 
three-dimensional universe models of modern relativistic cosmology. The main 
goal of the proposal is the appropriate appreciation of the intrinsic 
relationship between Hubble's law and the homogeneity of expanding 
relativistic models.  
\end{abstract}

\bigskip
\bigskip

\section{Introduction}
The first relativistic models of the universe appeared soon after the publication, in 1915, 
of the final formulation of the General Relativity Theory (GRT). In 1917, Albert Einstein 
and Willem de Sitter presented their models and in the beginning of the 1920s 
Alexander Friedmann presented his \cite{so12,viso}. All of these models are solutions of 
the field equations of GRT for idealized universes, namely, homogeneous universes.  Homogeneity simplifies enormously the  form of the field equations \cite{so13}. 
The de Sitter and Friedmann solutions represent expanding universes (or, contracting, for 
one of Friedmann's models).  

The relativistic models of the universe started to have a greater impact in the scientific 
community with the advent of astronomical observations, which were consistent with the 
fundamental feature of theoretical models, i.e., their spatial homogeneity.  Such 
observations were the result of the work of many astronomers but were synthesized and 
presented in a convincing way by one of them, the American astronomer Edwin Hubble, 
in the form of a relation that became known as ``Hubble's law''. It indicates that galaxies 
are receding from each other in such a way that the greater the distance between them  
the greater their recession speeds.  Mathematically, the Hubble law of expansion is 
written as  v $\propto$ d, where v is the galaxy recession speed and d is its distance to a 
point of reference, namely, the observer location on Earth.  
 
The connection between the homogeneity of relativistic models and Hubble's law of  
expansion is not always properly appreciated in presentations of modern relativistic 
cosmology. In general, it is only stressed the existence of the expansion itself and of its 
more immediate consequence, that is, the necessity of the existence of a beginning for the 
universe, derived from the extrapolation of the expansion to the past. The initial event is 
often called {\it Big Bang} or {\it initial singularity}. 

Now, the homogeneity of any expanding body is only preserved if the expansion  occurs 
according to a law of the type {\it velocity of expansion proportional to distance}.
In other words, all points of the body must have velocities of expansion proportional 
to their distances to an arbitrary  reference point, which is also located in the body. 
Homogeneity results from the fact that each part of the body expands equally along 
the expanding body. The uniform expansion of parts does not destroy the homogeneity of 
the whole. It is the cumulative effect of such small increments along the body that gives rise to 
a law of expansion of the type velocity proportional to distance. 

The body in question here is the universe. If it expands and stays homogeneous, then 
its expansion must obey a law like the one described above. And that is precisely what 
Edwin Hubble showed in a clean and objective way at the end of the 1920s. 

In the next section I discuss briefly Hubble's law, presenting the meaning of the terms that 
appear in its formulation. I show in the third section an unidimensional analogue of the 
three-dimensional expanding models, which illustrates, in a practical and very simple way,  
the intrinsic relation between homogeneity and a law of the type v $\propto$ d.  In the final section 
I make some considerations about the possible perceptions of the universe, which can be, in 
general, theoretical and observational. The English cosmologist Edward Milne called these 
perceptions ``world map'' and ``world picture'', respectively.  The unidimensional analogue 
put forward here is considered in this context and helps the understanding of such 
generalizations of universe perceptions.  

\section{Hubble's law}
Hubble's law is the observational foundation of a theoretical proposition, based in the GRT, 
known as ``the expansion of the universe''. According to this proposition, the universe is expanding 
in such a way that galaxies are receding faster the further away they are.  The 
discovery of Hubble's law \cite{hu29} was one of the greatest scientific achievements   
of the 20th century.  A detailed description of the events and scientists that contributed 
for its discovery is presented in \cite[chap. 14]{harr}. 

\begin{figure}[H]
\begin{center}
\includegraphics[width=10cm]{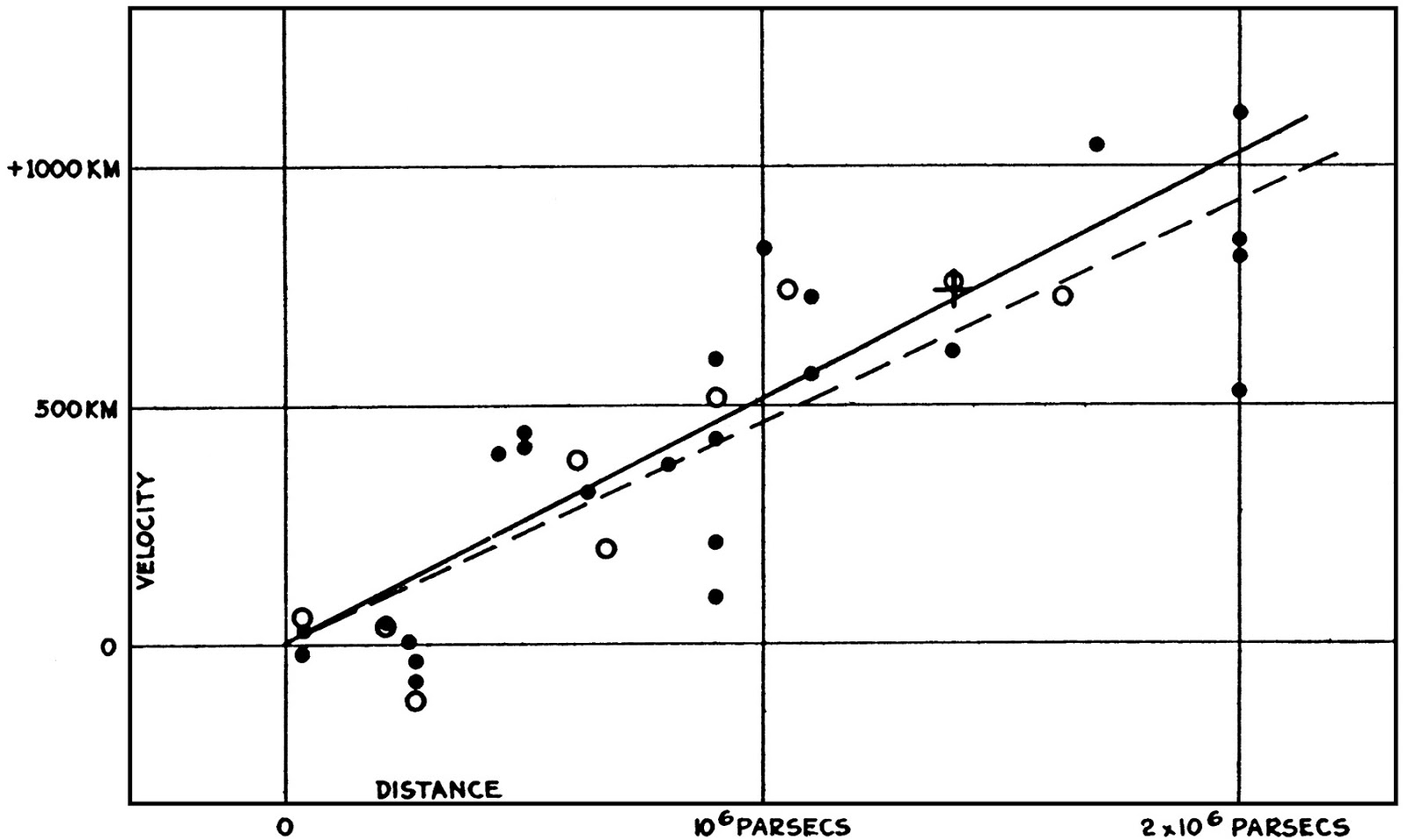}
\end{center}
\caption{Hubble's law v=\Ho d in the way it was presented in 1929 \cite[Fig. 1]{hu29}. The 
continuous and dashed lines represent fitting to the data according to distinct 
sampling criteria. Filled circles (continuous line) represent 24 individual galaxies and circles 
9 groupings of the same galaxies. Distances, in the abscissa axis, are in parsec (1 pc = 
3.26 light-year) and velocities in km/s. }
\label{fig:hu29}
\end{figure}

Fig. \ref{fig:hu29} is a reproduction of the diagram, presented by Hubble in his 1929 article, which 
is the graphical representation of Hubble's law. There one sees two lines that represent different 
fittings to the data of velocities and distances of galaxies outside the Local Group of galaxies. The 
continuous line is the fitting to the filled circles, which represent data of 24 galaxies. The dashed line 
is the fitting to the circles, which represent groupings of the 24 galaxies in 9 groups, according  
to the galaxy proximity in distances and on the sky plane. The fitting for both lines --- with different 
values of their slopes --- is  
\medskip 
\begin{equation}
\label{eq:eq1}
v=H_\circ d,
\end{equation}
which is Hubble's law, where \Ho\ is called 
``Hubble's constant''.  Velocities were obtained from the galaxy spectra. The spectral lines present 
in the spectra are displaced to wavelengths that are systematically larger than the same spectral lines 
measured in the laboratory. Such displacement --- called ``redshift'' --- may be interpreted as 
due to the motion of recession of the galaxies with respect to the observer. In the same way, an 
approaching motion results in a blueshift of the spectral lines (more details in \cite[chap. 14]{harr}).

\bigskip\bigskip\bigskip
\bigskip

The redshift is usually represented by the letter z and is defined as 
z $\equiv \Delta\lambda/\lambda$, with $\Delta\lambda=\lambda_\circ-\lambda$, where  
$\lambda_\circ$ is the observed wavelength of a given spectral line and $\lambda$ is the 
wavelength of the same line measured in the laboratory on Earth. Redshifts can be transformed 
into recession velocities by means of the mathematical expression of the classical Doppler 
effect v = cz, where z is the redshift and c is the speed of light in vacuum. 

Distances to the galaxies were determined by various methods, all taking into account the 
dependence of the observed luminous flux with the inverse of the squared distance to the  
observed object. In most of the cases, Hubble determined the luminous flux of individual stars 
present in galaxies and from a comparison with the luminous flux of similar stars in our 
Milky Way, with known distances, he calculated the distances to the host galaxies of the observed 
stars. In some cases he used the observed luminous flux of a whole galaxy and compared it with 
the flux of closer similar galaxies, whose distances were already known from other methods.  
Distance determinations were precarious, but constituted the best one could do in the end of the 
1920s. The Hubble constant \Ho\ determined by Hubble was almost 10 times as large as its value 
known in present days, mainly because of the uncertainty in distances.  Nevertheless, at that 
time,  the important fact was the convincing establishment of the linear relation between v and d. 

The fundamental importance of Hubble's law is that a law of this kind represents a necessary 
feature of the homogeneity of expanding or contracting universe models. The more common cases, therefore the ones we are interested on, are expanding models.These models are only obtained  
when the assumption of homogeneity of space is made in the solution of GRT's field equations (see 
detailed discussion in \cite{so13}). If a law of this kind is observationally verified in the real 
universe, then it means that the assumption that the universe might be spatially homogeneous 
has an observational foundation and is not a mere theoretical assumption. In the following 
section I show that a law of the type v $\propto$ d is indeed a consequence of spatial 
homogeneity by means of the investigation of a simple mechanical analogue of the expanding 
universe.    

\section{The expanding rubber band}
\begin{figure}[H]
\begin{center}
\includegraphics[width=8cm]{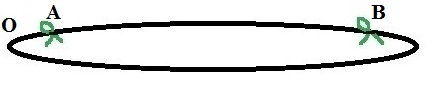}
\end{center}
\caption{The string loops tied to the rubber band represent ``galaxies'' {\bf A} and {\bf B}. 
For the purpose of discussion, the observer {\bf O} is located at the left end of the band. 
He may however be on any point of the ERBU.}  
\label{fig:uge1}
\end{figure}

Here I introduce ERBU, the {\bf Expanding Rubber Band Universe}. ERBU is a 
homogeneous rubber thread shaped in a closed figure, as illustrated in Fig. \ref{fig:uge1}. 
ERBU is the unidimensional 
equivalent to ERSU,  Edward Harrison's {\bf Expanding Rubber Sheet Universe} \cite[pp. 275 
to 280]{harr}, which is a 2-D analogue to the 3-D expanding universe.  As I show below, ERBU 
is much more practical  than ERSU --- because it is 1-D and of easy construction --- to be used 
in a demonstration of Hubble's law in the study of relativistic cosmology. An unidimensional analogue  
like this has already been discussed in the context of Hubble's law by Bernard Schutz 
(cf. \cite[p. 349]{schu}, where it is called {\it Rubber-Band Model of the Universe}, RBMU). RBMU is 
used by this author for quantitative applications of Hubble's law \cite[p. 348]{schu}, and the aspect 
I discuss here, namely, the homogeneity of relativistic models of the universe, is mentioned 
but is not appropriately highlighted, specially the straight relation between Hubble's law and 
homogeneity. Another difference between ERBU and RBMU is that in the first one we treat expansion 
(the stretching of the rubber band) in only one direction, i.e., a linear expansion. In RBMU the 
rubber band expands from an initial circular shape and must keep this shape.  The linear 
expansion is less rich in detail but is sufficient to the objective of the present work. RBMU's study is 
an important complement to the study presented here.  
 
Fig. \ref{fig:uge1}  shows ERBU. The material used in its confection is constituted by 
a rubber band and two pieces of string.

\bigskip\bigskip\bigskip
\bigskip

The ERBU of Fig. \ref{fig:uge1} may be used in three different ways. (i) Fixing the left end {\bf O} 
and stretching the right end one notes that the left loop {\bf A} moves much less than the right 
loop {\bf B}. This is a behavior that follows from the ERBU Hubble's law: the larger the distance to 
the ``observer'', in this case the left end {\bf O}, the larger the galaxy's displacement and 
therefore the larger its speed. In this way, v is  proportional  to the distance d, as explained in 
the previous section. (ii) Fixing the right end and stretching the left end, the loops behavior is 
reversed; the speed of galaxy {\bf B} is now lesser than the speed of galaxy {\bf A}. (iii) 
Simultaneously moving both ends, the previous scenarios happen simultaneously too; in other 
words, there is no privileged observer.  

\begin{figure}[H]
\begin{center}
\includegraphics[width=8cm]{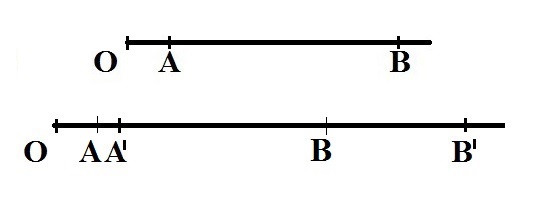}
\end{center}
\caption{The lines represent the rubber band of Fig. \ref{fig:uge1}. The distance from the 
observer to galaxy {\bf B} is, in this example, equal to 7 times the distance to galaxy 
{\bf A}. The bottom line shows the expanded rubber band. The band is homogeneous 
and remains homogeneous after the expansion and hence one has {\bf OB'}=
7$\times${\bf OA'}. The fine traces in the bottom line denote the initial positions of the galaxies.}
\label{fig:uge2}
\end{figure}

All that occurs in order to preserve the band homogeneity. Fig. \ref{fig:uge2} shows 
the ERBU of Fig. \ref{fig:uge1} before and after the expansion --- or stretching --- of the rubber 
band.

Let us assume that the band is stretched during a time interval $t_\circ$. The displacement of 
galaxy {\bf A} in this interval is {\bf AA'}. One may imagine that such displacement is the sum of 
small displacements which occur along {\bf OA}. Since the band is homogeneous, all these small 
displacements have the same magnitude $\delta$d. The total displacement {\bf AA'} is equal to 
$N\times\delta$d, where $N$ is the number of displacements $\delta$d. Hence, the larger is {\bf OA}, 
the larger is the number $N$ and, consequently, the larger is {\bf AA'}, that is to say, {\bf AA'} is 
proportional to {\bf OA}. Likewise, the displacement {\bf BB'}, of galaxy {\bf B}, is 
proportional to {\bf OB}. Such behavior patterns result in a ``Hubble's law'' for the ERBU, as 
shown next. 

\bigskip\bigskip\bigskip
\bigskip

Let $\gamma$ be the constant of the proportionality described above.  One has then 
d'=$\gamma$d, where d' represents the generic displacement of a galaxy from its initial distance 
to the reference point {\bf O}. Note that $\gamma$  is a dimensionless constant. For the galaxy 
{\bf A}, for example, d'={\bf AA'} and d={\bf OA}. Since the displacement d' occurred during the 
time interval $t_\circ$, one can write then  d'=v$t_\circ$, where v is the galaxy velocity in the 
displacement. In Fig. \ref{fig:uge2} one can see that that the displacement velocity of galaxy {\bf B} 
will be larger than the displacement velocity of galaxy {\bf A}, because in the same time interval 
$t_\circ$ its displacement was {\bf BB'}=7$\times${\bf AA'}. The velocity of {\bf B} will be 
therefore 7 times as large as the velocity of {\bf A}. 

The relation d'=$\gamma$d becomes hence v$t_\circ = \gamma$d, or v=($\gamma/t_\circ$)d. 
Making $\gamma/t_\circ \equiv$  H$_G$, the {\it Hubble's constant} of the ERBU, we have the 
expression of Hubble's law for the ERBU as: 
\medskip
\begin{equation}
\label{eq:eq2}
v=H_Gd,
\end{equation}
where v is the velocity of displacement of any point of the rubber band when the band is stretched, 
H$_G\equiv\gamma/t_\circ$ is the stretching --- or expansion --- constant of the rubber band and d 
is the distance of the point to the reference  {\bf O}. Eq. \ref{eq:eq2} is entirely analogous to 
eq. \ref{eq:eq1} of Hubble's cosmological expansion, and one may note in both equations that 
\Ho\ and H$_G$ have physical dimensions of 1/time. 

In the expanding universe models, similarly to what occurs in the rubber band,  
Hubble's law describes an expansion that preserves the universe homogeneity and, as in the band, 
there is no privileged observer or point of reference as well.

The rubber-band ``Hubble's constant'' is related to its elasticity because a ``hard'' rubber band 
expands (or stretches) with more difficulty than a ``soft'' one. By analogy, one may say that 
the cosmological Hubble's constant is related to the {\it elasticity of the spatial tissue}. Space and space-time in GRT are physical entities. Rubber analogies of the universe, like the one presented 
here, show that it is conceptually appropriate attributing an elastic property to space  (space 
elasticity is discussed, for example, in \cite[pp. 286-287]{tawh}). 

It must be pointed out that Harrison's ERSU, being a rubber sheet, i.e., a 2-D object, 
has an important advantage over ERBU, namely, to enable the discussion of the behavior of 
two-dimensional patterns in expanding models  (e.g., \cite[p. 276]{harr}). Now,  
Schutz's RBMU expands keeping a circular shape, while ERBU expands linearly, a feature that is 
sufficient for the discussion of homogeneity, and it is also, because of this same feature, more easily 
handled  than RBMU is. The latter, as said before, needs to keep a circular shape while being 
stretched. 

The ERBU allows, therefore, the experimental verification of one of the most important 
consequences of the homogeneity of a physical system, namely, the validity of a law of the kind 
v $\propto$ d, in other words, the validity of a  ``Hubble's law''.

\section{Final remarks}
As we saw, ERBU is a simple mechanical analogue of the expanding universe, which allows  a better 
conceptual understanding of Hubble's law. It is suitable also for the discussion of two interesting 
cosmological concepts introduced by the English physicist, mathematician and cosmologist Edward 
Milne (1896-1950).  These are the {\it ``world map''} and the {\it ``world picture''}. They are 
general concepts and may be applied to any cosmological models, either in expansion or not.  The 
map and the picture of the universe are two possible ways of perception of the universe. Such 
concepts are explored more extensively in \cite[cap. 14]{harr}; I make next a brief presentation of 
their meanings. 

The {\it world map} is what is  
perceived by cosmic observes external to the universe, i.e., by {\it godlike spectators} 
(cf. \cite[p. 279]{harr}).  Such a spectator sees all the cosmos as it is in a given instant of 
time. In our analogue, the external spectator sees the whole ERBU, that is, the rubber band.  The 
external spectator is, generally speaking, everyone that handles the ERBU.   

As to the putative observer located in the ERBU's point {\bf O}, he sees, 
according to Milne, the ``world picture'', and has observational limitations that does not exist for 
the godlike observer. The observer located in {\bf O} is a {\it wormlike denizen} of the ERBU. 
For the real universe, such limitations are more obvious. There, the wormlike denizen sees bodies 
that are distant in space and remote in time and is unable of perceiving the whole cosmos as it is 
in {\it a given instant of time}, i.e., the world map, because of the finiteness of the speed of light. 
In the real universe this is, however, the only way of observing the universe. In other words, we 
are wormlike denizens of the real universe.  

The world  map is perceived by someone from the outside, being necessarily a theoretical view.  
The world picture is perceived by someone from the inside, being then an observational view. The 
perception of the expansion, in this case, has two fundamental limitations, one  arising from 
its observational nature, namely, the finiteness of the speed of light, and another of a theoretical 
nature, i.e., the necessity of choosing a cosmological model that describes the expansion. 
This choice determines the way redshifts are transformed into velocities, as well as 
how distances to observed galaxies 
are calculated.  In this way, the velocity-distance relation is only linear for small redshifts (and 
consequently small distances), because then the influence of the finiteness of the speed of light 
and of the adopted theoretical model for the expansion are negligible. For large distances the 
velocity-distance relation depends on the adopted model of expansion, because primary 
observational data are not velocities but redshifts. And these must be transformed into velocities 
in accordance with the model. The function v = cz, equivalent to the classical Doppler effect, is only 
valid for small redshifts z (like those used by Hubble in 1929); for large values of z, the function 
v(z) depends on the adopted model for the expanding universe. For example, for the critical 
Friedmann model  \cite{viso} this function is not linear and is illustrated in figure 2 of \cite{so09} 
and in figure 15.8 of \cite{harr}.  
 
In conclusion, the law v=\Ho d holds for any distance 
in the world map, as long as it is homogeneous, because as we saw in the previous section, 
it is a law of this sort that describes the homogeneity of expanding universes. However, for large 
distances in the world picture, the velocity-distance law is not, in general, linear. Non linearity 
starts to be observed for redshifts larger than 0.1 \cite{harr,so09}, that is,  for distances larger 
than about 1 billion light-years, for Hubble's constant of 72 (km/s)/Mpc  
\cite{freed} (1 Mpc $ = 3,26\times10^6$ light-year).

\bigskip
\bigskip

{\noindent \it Acknowledgments --- } I thank the comments of  the anonymous referee, 
which have contributed for a better presentation of the manuscript, and for making me aware 
of Bernard Schutz's rubber band model (cf. \cite[p. 349]{schu}).


\begin{thebibliography}{}
\bibitem{so12} D. Soares, {\it Einstein's static universe}  (2012), arXiv:1203.4513v2 
[physics.gen-ph]
\bibitem{viso} A. Viglioni e D. Soares, {\it Note on the classical solutions of Friedmann's 
equation} (2011),  arXiv:1007.0598 [physics.gen-ph]
\bibitem{so13} D. Soares,  {\it Physico-mathematical foundations of relativistic cosmology} 
(2013),  arXiv:1309.2252 [physics.gen-ph] 
\bibitem{hu29} Hubble, E.: A relation between distance and 
radial velocity among extra-galactic nebulae. Proceedings of the 
National Academy of Sciences {\bf 15}, 168 (1929).
\bibitem{harr} E. Harrison, {\it Cosmology -- The Science of the 
Universe}, (Cambridge University Press, Cambridge, 2000). 
\bibitem{schu} B. Schutz, {\it Gravity from the ground up}, (Cambridge University Press, 
Cambridge, 2003). (Available also at  \\
{\tt http://pt.slideshare.net/PedroPrez19/physics-gravity-from-the-\\ground-up-bernard-schutz-cambridge-university-press-2003}).    
\bibitem{tawh} E.F. Taylor, J.A. Wheeler, {\it Spacetime Physics, Introduction to Special 
Relativity}, (W.H. Freeman and Company, New York, 1992). 
\bibitem{so09} D. Soares, {\it The age of the universe, the Hubble constant and the 
accelerated expansion and the Hubble effect} (2009), arXiv:0908.1864 [physics.gen-ph]
\bibitem{freed} W.L. Freedman, B.F. Madore, B.K. Gibson, L. Ferrarese, D.D. Kelson, S. Sakai, 
J.R. Mould, R.C. Kennicutt, Jr., H.C. Ford, J.A. Graham, J.P. Huchra, S.M.G. Hughes, 
G.D. Illingworth, L.M. Macri e  P.B. Stetson,  Astrophysical Journal {\bf 553}, 47 (2001).
\end{thebibliography}
\end{document}